\documentclass[12pt,letterpaper]{article}
\usepackage[margin=1in]{geometry}
\usepackage{graphicx,amsmath,amssymb,amsfonts,bm,nicefrac,titling,cite} 

\begin{document} 

\title{Stability of Boolean networks: The joint effects of topology and update rules} 

\author{{Shane~Squires}, {Andrew~Pomerance}, {Michelle~Girvan}, and {Edward~Ott}\\[5pt]
\textit{University of Maryland, College Park, MD, USA}}

\date{\today} 

\maketitle 

\abstract{We study the stability of orbits in large Boolean 
networks with given complex topology.  We impose no restrictions 
on the form of the update rules, which may be correlated with 
local topological properties of the network.  While recent past 
work has addressed the separate effects of nontrivial network 
topology and certain special classes of update rules on stability, 
only crude results exist about how these effects interact.  We 
present a widely applicable solution to this problem.  Numerical 
experiments confirm our theory and show that local correlations 
between topology and update rules can have profound effects on 
the qualitative behavior of these systems.\\[6pt]
\textsc{pacs:} \texttt{89.75.-k} (complex systems), \texttt{05.45.-a}
(nonlinear dynamical systems), \texttt{64.60.aq} (phase transitions in networks).}

\section*{Introduction} 
\indent \indent
Systems formed by interconnecting collections of Boolean elements have 
been successfully used to model the macroscopic behavior of a wide variety 
of complex systems.  Examples include genetic control \cite{kauffman69}, 
neural networks \cite{hopfield82}, ferromagnetism \cite{glauber63}, 
infectious disease spread \cite{newman02-1}, opinion dynamics \cite{% 
castellano09}, and applications in economics and geoscience \cite{coluzzi11}.  
Each of these diverse models share the same basic structure: a set of 
nodes, each of which has a binary state ($0$ or $1$) at a given integer 
time $t$, and a set of update rules that determines the state of each 
node at time $t+1$ given the states of the nodes at time $t$.  The 
relationships between nodes define a graph, where an edge is drawn 
from node $j$ to node $i$ if the update rule for node $i$ depends on 
the state of node $j$.  

Depending on the desired application, the model's graph can be random 
\cite{kauffman69}, fully connected \cite{hopfield82}, a lattice \cite{% 
glauber63}, or have other complex topology \cite{kauffman03,pomerance09}.  
The states of nodes can be updated deterministically or stochastically, 
synchronously or asynchronously \cite{kauffman69,glauber63,hopfield82}.  
Finally, in cases where the update rules are considered to be randomly 
generated, they can be drawn from many different ensembles \cite{derrida86,% 
kauffman03,pomerance12,rohlf02}.  

One important question about a Boolean network is whether or not it is 
stable, i.e., whether or not small perturbations of a typical initial 
state tend to grow or shrink as the system evolves.  This question may 
have important ramifications for systems biology and neuroscience: it 
has been hypothesized that both gene networks \cite{kauffman93} and 
neural networks \cite{shew09} exist near the critical border separating 
the stable and unstable regimes. Recently, Pomerance et al.\ \cite{pomerance09} 
introduced the additional hypothesis that orbital stability of the gene 
regulatory system may be causally related to cancer.  Specifically, 
motivated by microdissection experiments showing genetic heterogeneity 
in tumors \cite{gonzalez-garcia02}, they suggested that mutations that 
promote instability may be a contributing factor for some types of cancers.  

In this paper, we present and numerically verify a general method for 
studying the stability of large, directed Boolean networks with locally 
tree-like topology.  Here, by a locally tree-like network we mean that, 
if two nodes $j$ and $i$ are connected by a short directed path from 
$j$ to $i$, it is very unlikely that there will exist a second such 
path of the same length.  This allows us to make the approximation 
that two inputs to a node are uncorrelated.  Analyses based on this 
approximation have been found to yield accurate results, even in cases 
where the network contains significant clustering \cite{pomerance09,% 
melnik11}, while making an analytic treatment of the system tractable.  
Our results offer a means of assessing the stability of a wide variety 
of Boolean network systems for which, up to now, no generally effective 
method has been available.  We demonstrate the general utility of our 
approach with two examples illustrating that the joint effects of network 
topology and update rules can have profound effects on Boolean network 
dynamics, which cannot be captured by previous theories.

\section*{Model} 
\indent \indent
Boolean networks are discrete-state dynamical systems in which each of 
the $N$ nodes of a network has a binary state $x_i(t)=0$ or $1$ at each 
integer-valued time $t$, and is updated at the next time $t+1$ to a new 
binary state $x_i(t+1)$ that is determined from the time $t$ states of 
its network inputs.  For now we assume that updates are synchronous, 
but in the Supporting Information ({\em SI}), we demonstrate 
that the stability criterion that we obtain is the same 
whether nodes are updated synchronously or asynchronously.  Consider 
a node $i$ which has $K_i$ network inputs, $j_1$\dots$j_{K_i}$.  The 
new state of node $i$ is determined by a binary-valued update rule 
$F_i$, according to $x_i(t+1) = F_i\left( x_{j_1}(t),\dots,x_{j_{K_i}}(t) 
\right)$. Each $F_i$ may be specified in the form of a ``truth table'' 
listing all the $2^{K_i}$ possible inputs and the corresponding output.  
Denoting the vector of input states to node $i$ at time $t$ as $X_i(t) 
= (x_{j_1}(t), \dots, x_{j_{K_i}}(t))$, we have 
\begin{equation} 
\label{eq.dynamics2} 
x_i(t+1) = F_i \left( X_i(t) \right). \\ 
\end{equation} 
The network structure is represented using an adjacency matrix $A$, 
where $A_{ij} = 1$ if there is an edge $j \to i$ [that is, if $x_i(t+1)$ 
depends on $x_j(t)$], and $A_{ij} = 0$ otherwise.  

The question we address is whether the dynamics resulting from Eq.~% 
\eqref{eq.dynamics2} are stable to small perturbations.  To define 
stability, we assume $N \gg 1$ and consider two states, $\bm{x}(t) = 
( x_1(t), \dots, x_N(t) )^T$ and ${\tilde{\bm{x}}(t)} = ( \tilde{x}_1(t), 
\dots, \tilde{x}_N(t) )^T$.  We define the normalized Hamming distance 
between these two states as the fraction of the nodal values that differ 
for the two states, 
\begin{equation} 
H(\bm{x}(t) , \tilde{\bm{x}}(t)) = \frac{1}{N} \sum_{i=1}^N 
\left\vert x_i(t) - \tilde{x}_i(t) \right\vert. 
\end{equation} 
We consider $\tilde{\bm{x}}(0)$ to be a state that is slightly perturbed 
from $\bm{x}(0)$, meaning that $H(\bm{x}(0) , \tilde{\bm{x}}(0)) \ll 1$.  
Stability is then defined by whether $H(\bm{x}(t), \tilde{\bm{x}}(t))$ 
decreases to zero or grows to ${\mathcal O}(1)$ as $\bm{x}(t)$ and 
$\tilde{\bm{x}}(t)$ evolve under Eq.~\eqref{eq.dynamics2}.  Our main 
theoretical result is a criterion for stability that accounts for the 
joint effects of network topology (i.e., the $A_{ij}$) and node dynamics 
(i.e., the functions $F_i$). 

The stability of Boolean networks was addressed in the original 
work of Kauffman \cite{kauffman69,kauffman93}, where Boolean 
networks were first proposed as a model for genetic dynamics.  
Kauffman assumed a so-called $N$-$K$ network topology in which 
$K_i$ was the same at each node, $K_i=K$, and the $K$ inputs to 
each node were chosen randomly from amongst the $(N-1)$ other 
nodes.  Further, for each of the $2^K$ input states, $F_i$ was 
chosen to be $0$ or $1$ with probability $\nicefrac{1}{2}$.  
Derrida and Pomeau \cite{derrida86} later generalized this model 
by introducing a truth table bias $0<\bar{p}<1$ such that, for 
a given input, $F_i = 1$ with probability $\bar{p}$.  They also 
proposed a method of stability analysis for the case of ``annealed'' 
systems, described as follows.  First, note that the problem that 
they and Kauffman were interested in was one in which the network 
($A_{ij}$) and the node dynamics ($F_i$) are initially randomly 
chosen, fixed forever after, and then used to create the dynamics 
(``quenched randomness'').  The annealed problem is different in 
that the random choices of the network and node dynamics are made 
{\em at every time step}.  In contrast with the stability of the 
quenched system, the stability of the annealed system can be 
analytically determined \cite{derrida86}.  Derrida and Pomeau 
conjectured that, in the large $N$ limit, the stability boundaries 
for the quenched and annealed situations are approximately the same.  
This conjecture has been very well supported by the results of 
numerical experiments.  Later authors generalized the Derrida-Pomeau 
annealing approach to include a distribution of in-degrees \cite{% 
sole94,luque97,fox01,aldana03,aldana03-1}, joint in-degree/out-degree 
distributions \cite{lee08}, and ``canalizing'' update rules \cite{% 
shmulevich04,kauffman03,kauffman04}. 

A further significant generalization was presented in Ref.~\cite{% 
pomerance09} in which the network is quenched (not annealed), but 
the update rules ($F_i$) are annealed using a truth table bias 
$\bar{p}_i$ that may vary from node to node.  Reference~\cite{% 
pomerance09} called this procedure ``semi-annealing'' and used it 
to study the effects of network topological properties on stability, 
including such factors as network degree assortativity, correlation 
between node degree and the node bias $\bar{p}_i$, and community 
structure.  As in the case of annealing, numerical results strongly 
support the hypothesis that the stability of the semi-annealed 
(analytically treatable) system and a typical quenched system are 
similar \cite{pomerance09,pomerance12}. 

In what follows, we generalize the results of Ref.~\cite{pomerance12} 
by using a semi-annealing procedure that enables the treatment of 
previously inaccessible cases of substantial interest in applications.  
We then illustrate this new procedure using two numerical examples.  
The first example is primarily pedagogical.  The second is more 
application-oriented and uses threshold rules of the form 
\begin{equation} 
\label{eq.threshold} 
x_i(t+1) = U {\bigg (} \sum_j w_{ij} x_j(t) - \theta_i {\bigg )}, 
\end{equation} 
where $U$ denotes the unit step function, $\theta_i$ is a threshold 
value, and $w_{ij}$ is a signed weight whose magnitude reflects the 
strength of the influence of node $j$ on node $i$ ($w_{ij}=0$ if 
$A_{ij}=0$) and whose sign indicates whether node $j$ ``activates'' 
or ``represses'' node $i$ (i.e., promotes $x_i$ to be 1 or 0).  
(This model has been considered previously in the case where the 
network is $N$-$K$, $\theta_i=0$, and $w_{ij}=\pm 1$ \cite{kurten88,% 
rohlf02,greil07,szejka08}.)  Threshold networks are also commonly used 
to model gene regulation \cite{li04,rybarsch12}, neural networks 
\cite{hopfield82}, and other applications. 

We begin by specifying our semi-annealing procedure, which is similar to 
the probabilistic Boolean networks described in \cite{shmulevich02}.  We 
assign each node $i$ an ensemble of update rules, ${\mathcal T}_i$, from 
which a specific update rule is randomly drawn at each time step $t$.  
This choice is made independently at each network node $i$, and we denote 
the probability of drawing update rule $f$ as $\Pr[F_i=f]$.  The resulting 
dynamics may be described by the probabilities $q_i(X_i)$ that the state 
of node $i$, given inputs $X_i$, will be $1$ on the next time step, 
\begin{equation} 
\label{eq.dynamics3} 
q_i(X_i) = \sum_{f \in {\mathcal T}_i} \Pr[F_i=f] f(X_i), 
\end{equation} 
where we have used the fact that $f\left(X_i\right)=0$ or $1$.  
It is important to note that $q_i(X_i)$ is solely determined from 
${\mathcal T}_i$, independent of the update rule assignments at 
other nodes.  Thus, computation of $q_i(X_i)$ is straightforward.  

The advantage of this semi-annealing procedure is that the resulting 
dynamics are simpler to analyze than those of systems with quenched 
update rules.  Typically, the semi-annealed dynamics described by 
Eq.~\eqref{eq.dynamics3} possess a single ergodic attractor, and the 
stability of this attractor is similar to that of typical attractors 
in quenched systems.  We assume the existence of a single ergodic 
attractor in our analysis below and briefly comment on cases for 
which this assumption fails in the {\em SI}. 

When using the semi-annealed model, the selection of deterministic 
update rules is replaced by the selection of an update rule ensemble 
for each node $i$.  The choice of ${\mathcal T}_i$, like the choice 
of $F_i$ in deterministic models, depends on the particular case 
under study.  This will be illustrated in our numerical experiments. 

To measure the stability of the semi-annealed dynamics generated 
by Eq.~\eqref{eq.dynamics3}, we begin with many initial conditions 
$\bm{x}(0)$ and generate orbits $\bm{x}(t)$.  We imagine that the 
initial conditions $\bm{x}(0)$ are selected randomly according to 
the natural measure of the attractor; in practice, this can be 
achieved by time-evolving another initial condition sufficiently 
long that transient behavior has ceased, and then using its final 
state as an initial condition.  For each orbit $\bm{x}(t)$, we 
also consider a perturbed initial condition $\bm{\tilde{x}}(0)$, 
obtained by randomly choosing a small fraction $\varepsilon$ of 
the components of $\bm{x}(0)$ and ``flipping'' their states.  
That is, if node $i$ is chosen to be flipped, then $\tilde{x}_i(0) 
= 1 - x_i(0)$.  The perturbed initial condition is then used to 
generate a perturbed orbit $\bm{\tilde{x}}(t)$, where, for the 
semi-annealed case, the random update rule time sequence for each 
node is the same for $\bm{\tilde{x}}(t)$ and $\bm{x}(t)$.  The 
growth or decay of the Hamming distance between $\bm{x}(t)$ and 
$\bm{\tilde{x}}(t)$ defines the stability of the system.

\section*{Analysis} 
\indent \indent
Given an orbit on the ergodic attractor of the semi-annealed system, 
we define $p_i$ to be the fraction of time that the state $x_i(t)$ 
of node $i$ is $1$.  We call $p_i$ the ``dynamical bias'' of node $i$ 
and regard it as the probability that $x_i(t)=1$ at an arbitrarily 
chosen time.\footnote{This is in contrast to the ``truth table bias,'' 
denoted $\bar{p}$ above, an external parameter used to define the 
ensemble of update rules in Ref.~\cite{derrida86} and later work.}  
In what follows, we first address the determination of the dynamical 
biases $p_i$, which can then be used to determine the stability of 
the system. 

We first note that $p_i$ is determined by the set of probabilities 
$\Pr[X_i]$ of $i$ receiving each input vector $X_i$, using 
$p_i = \sum_{X_i} \Pr\left[ x_i = 1 \left\vert X_i \right.\right] 
\Pr \left[ X_i \right]$, or 
\begin{equation} 
\label{eq.p} 
p_i = \sum_{X_i} q_i(X_i) \Pr\left[ X_i \right], 
\end{equation} 
where $q_i$ is as defined in Eq.~\eqref{eq.dynamics3}.  Assuming 
that the network topology is locally tree-like, the states of the 
inputs to node $i$ can be treated as statistically independent 
\cite{pomerance09,melnik11}. Therefore, the probabilities $\Pr[X_i]$ 
are determined by the biases of $i$'s inputs.  Letting ${\mathcal J}_i$ 
denote the set of indices of all nodes that are inputs to $i$, 
\begin{equation} 
\label{eq.PrX} 
\Pr\left[ X_i \right] = \prod_{j \in {\mathcal J}_i} \left[ x_j p_j + 
\left( 1 - x_j \right) \left( 1 - p_j \right) \right], 
\end{equation} 
where we have used the fact that $x_i=0$ or $1$.  Inserting \eqref{eq.PrX} 
into \eqref{eq.p} yields a set of $N$ equations for the $N$ node biases 
$p_i$.  In what follows, we envision that this set of equations has 
been solved for the dynamical biases $p_i$ at each node, and we will 
use these biases to evaluate the stability of the network.  We find that 
for most practical purposes, one method for solving Eqs.~(\ref{eq.p}--% 
\ref{eq.PrX}) for the biases $p_i$ is by iteration: an initial guess for 
each $p_i$ can be inserted in \eqref{eq.PrX}, and \eqref{eq.p} can then 
be used to obtain an improved guess, and so forth, until the $p_i$ have 
converged. 

We now consider the stability of the annealed system.  We say that node 
$i$ is ``damaged'' at time $t$ if $x_i(t)$ and $\tilde{x}_i(t)$ differ 
at time $t$.  We define a vector $\bm{y}(t)$ such that $y_i(t)$ is the 
probability that $i$ is damaged at time $t$, i.e., 
\begin{equation} 
\label{eq.y1} 
y_i(t) = \Pr\left[ x_i(t) \ne \tilde{x}_i(t) \right]. 
\end{equation} 
Next, let $d_i(X_i, \tilde{X}_i)$ be the probability that $i$ will be 
damaged if its inputs in the two orbits are $X_i$ and $\tilde{X}_i$, 
\begin{equation} 
\label{eq.d1} 
%Used manual spacing here to prevent equation from running into equation label. 
d_i\hspace{-2pt}\left( X_i, \tilde{X}_i \right)\hspace{-2pt}= 
\hspace{-2pt}\Pr\hspace{-2pt}\left[x_i(t+1) \ne \tilde{x}_i(t+1) 
\left\vert X_i(t),\tilde{X}_i(t) \right. \right]. 
\end{equation} 
We have suppressed the time dependence of $X_i$ and $\tilde{X}_i$ in 
$d_i( X_i, \tilde{X}_i)$ since this can be expressed in terms of the 
time-independent update rule ensemble as 
\begin{equation} 
\label{eq.d2} 
d_i\left( X_i, \tilde{X}_i \right) = \sum_{f \in {\mathcal T}_i} 
\Pr\left[ F_i=f \right] \cdot \left\vert f(X_i) - f(\tilde{X}_i) \right\vert, 
\end{equation} 
where we have used the fact that $f(X_i)=0$ or $1$.  Note that, 
like $q_i$, $d_i$ depends only on ${\mathcal T}_i$, and thus is 
straightforward to calculate.  

Marginalizing over $X_i$ and $\tilde{X}_i$ in \eqref{eq.y1} and 
inserting \eqref{eq.d1}, 
\begin{equation} 
\label{eq.y2} 
y_i(t+1) = \sum_{\phantom{\tilde{1}} X_i \phantom{\tilde{1}}} 
\sum_{\tilde{X}_i} \Pr\left[X_i(t), \tilde{X}_i(t)\right] d_i\left[ X_i, 
\tilde{X}_i \right]. 
\end{equation} 
Because we are considering the question of stability, we have assumed 
that $\bm{x}(t)$ and $\bm{\tilde{x}}(t)$ are close to each other in 
the sense of Hamming distance for small times $t$, so $y_i(t) \ll 1$ 
for all $i$.  In this case we can ignore the possibility that $X_i(t)$ 
and $\tilde{X}_i(t)$ differ at two or more input states and drop all 
terms of ${\mathcal O}(y^2)$.  Moreover, if $\tilde{X}_i(t)$ and $X_i(t)$ 
are the same, $d_i=0$ via Eq.~\eqref{eq.d2}, so nothing is contributed 
to the sum in Eq.~\eqref{eq.y2}.  Therefore, the only values of 
$\tilde{X}_i$ which contribute significantly to the sum are the ones 
in which $\tilde{X}_i(t)$ and $X_i(t)$ differ for exactly one node $j$.  
Let $X_i^j(t)$ be a vector which is the same as $X_i(t)$ except that 
the state of input node $j$ is flipped [$\tilde{x}_j(t)=1-x_j(t)$].  
Using this notation, we can rewrite Eq.~\eqref{eq.y2} as 
\begin{equation} 
\label{eq.y3} 
y_i(t+1) = \sum_{j \in {\mathcal J}_i} \sum_{X_i} \Pr\left[X_i(t), 
X_i^j(t) \right] d_i\left( X_i, X_i^j \right). 
\end{equation} 
Furthermore, because the network is locally tree-like and the inputs to 
node $i$ are therefore uncorrelated, 
\begin{equation} 
\begin{split} 
\Pr\left[X_i(t), X_i^j(t) \right] &= \Pr\left[X_i\right] \Pr\left[x_i(t) 
\ne \tilde{x}_j(t)\right] \\ 
&= \Pr\left[X_i\right] y_j(t). 
\end{split} 
\end{equation} 
When substituted into Eq.~\eqref{eq.y3}, this leads to 
\begin{equation} 
y_i(t+1) = \sum_{j \in {\mathcal J}_i} y_j(t) \sum_{X_i} \Pr\left[X_i\right] 
d_i\left( X_i, X_i^j \right). 
\end{equation} 
Since the second sum is time-independent, we can write 
\begin{subequations} 
\label{eq.r} 
\begin{align} 
y_i(t+1) &= \sum_j R_{ij} y_j(t) + {\mathcal O}(y^2), \\ 
R_{ij} &\equiv \sum_{X_i} \Pr\left[X_i\right] d_i\left( X_i, X_i^j \right), 
\end{align} 
\end{subequations} 
where $R_{ij}=0$ when there is no edge from $j$ to $i$.\footnote{The 
second-order terms in this expansion are discussed further in the {\em 
SI}, where we derive an expression for the critical slope of 
the stability phase transition.}  $R_{ij}$ may be interpreted as the 
probability that damage will spread from node $j$ to node $i$; in 
analogy with the terminology of Ref.~\cite{shmulevich04}, we call 
it the effective activity of $j$ on $i$.  

The average of the normalized Hamming distance over all possible 
perturbations and realizations of the semi-annealed dynamics is 
$\langle H(\bm{x}(t), \bm{\tilde{x}}(t)) \rangle = \tfrac{1}{N} 
\sum_i y_i(t)$, so the stability of the system is determined by 
whether or not the elements of $\bm{y}$ grow with time.  This can 
be determined by writing Eq.~(\ref{eq.r}a) in matrix form, 
\begin{equation} 
\label{eq.yt} 
\bm{y}(t+1) = \bm{R} \hspace{1pt} \bm{y}(t). 
\end{equation} 
Since the effective activities $R_{ij}$ are non-negative, and 
$\bm{R}$ is typically a primitive matrix, the Frobenius-Perron 
theorem implies that the eigenvalue of $\bm{R}$ with largest 
magnitude is real and positive.  We denote this eigenvalue 
$\lambda_R$.  If the initial perturbation has a nonzero component 
along the eigenvector associated with $\lambda_R$, as is generally 
the case, then for $t$ not too large, the expected Hamming distance 
will grow as $(\lambda_R)^t$ by Eq.~\eqref{eq.yt}.  Therefore, 
\begin{equation} 
\label{eq.stability} 
\left. \begin{matrix} \lambda_R > 1 \text{ implies instability} \\ 
\lambda_R < 1 \text{ implies stability\phantom{in}} 
\end{matrix} \right\}. 
\end{equation} 

One major advantage of our analysis is that, from a computational 
perspective, evaluating $\lambda_R$ is typically faster than finding 
the average Hamming distance through simulations. We discuss this 
further in the {\em SI}, along with other computational 
aspects of the above solution.  Another potential advantage of the 
criterion \eqref{eq.stability} is that, in some cases, it can 
facilitate qualitative understanding.  For example, in previous 
work \cite{pomerance09}, it was shown that network assortativity 
promotes instability for a special case of the above situation.

\section*{Numerical results}  
\begin{figure*}[t] 
\centering 
\includegraphics{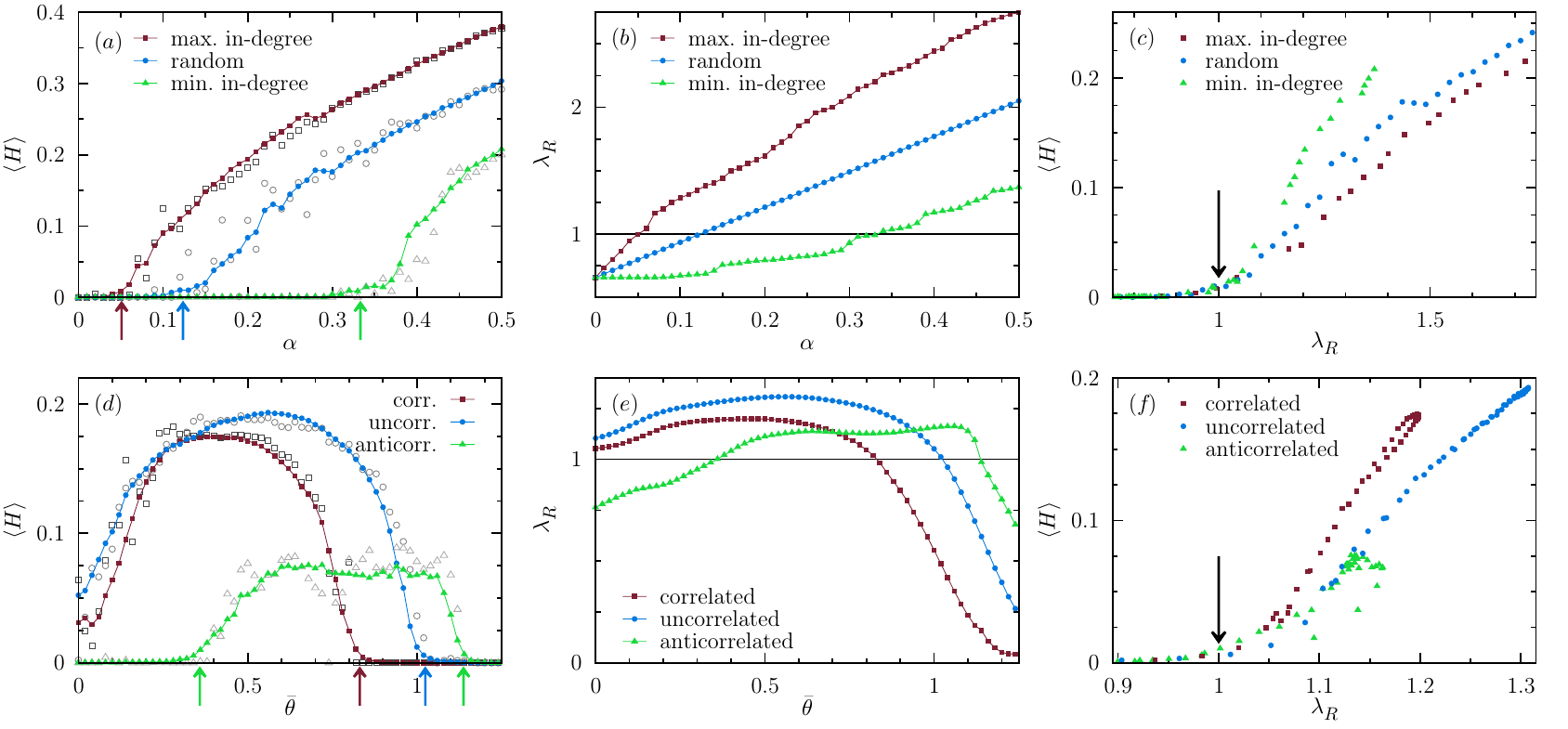}
\caption{Normalized average Hamming distance $\langle H \rangle$ and 
$\lambda_R$ for a network with {\small XOR}, {\small OR}, and {\small 
AND} update rules (panels $a$--$c$) and a threshold network (panels 
$d$--$f$).  Filled markers are averaged over $50$ quenched realizations 
of the thresholds, while hollow markers show a single quenched 
realization.  Squares, circles, and triangles represent different 
correlations between network topology and update rules; see text for 
details.  ($a$) When $\alpha$ is used as a tuning parameter, the 
stability transitions for each of the three cases are far apart.  
Locations where $\lambda_R=1$ are marked by arrows.  ($b$) Viewing 
$\lambda_R$ as a function of $\alpha$, we see that this behavior 
agrees with the theoretical prediction for critical stability, 
$\lambda_R=1$.  ($c$) Plotting $\langle H \rangle$ against $\lambda_R$ 
directly, we see that the transition for all three curves occurs 
at $\lambda_R=1$.  ($d$--$f$) Results for threshold networks are 
shown, analogous to those in panels ($a$--$c$), using $\bar{\theta}$ 
as a tuning parameter rather than $\alpha$.} 
\label{fig} 
\end{figure*}

\indent \indent
We now use the general framework presented above to analyze two 
cases that illustrate the effects of correlations between local 
topological features and update rules.  In each example, we 
construct a single network with $N=10^5$ nodes using the 
configuration model \cite{newman03}.  The in-degrees are Poisson-% 
distributed with a mean of $4$ and the out-degrees are scale-free 
with exponent $\gamma \approx 2.2$.  In Fig.~\ref{fig}, we plot 
the average Hamming distance $\langle H \rangle$ and $\lambda_R$ 
against a tuning parameter for each model.  To calculate each 
Hamming distance $H$, we first time-evolve a random initial 
condition (using a quenched set of update rules) for $100$ time 
steps to ensure that it is on an attractor, then apply a perturbation 
by flipping the values of a random fraction $\varepsilon=0.01$ of 
the nodes.  Next we time-evolve both the original and perturbed 
orbits for another $400$ time steps, measuring the Hamming distance 
$H$ over the last $100$ of these to ensure that it has reached a 
steady state.  We take the average $\langle H \rangle$ over $10$ 
initial conditions for each quenched set of update rules.  In the 
figures, we show $\langle H \rangle$ for both a single quenched 
system as well as an average over $50$ sets of quenched update rules. 

\subsection*{Example 1: {\small XOR}, {\small OR}, and {\small AND} 
            update rules}  
\indent \indent
In our first example, we illustrate the effect of correlations by 
assigning a node either a highly ``sensitive'' update rule ({\small 
XOR}) or a less sensitive update rule ({\small OR} or {\small AND}) 
based on the node's in-degree.  That is, the update rule at each 
node $i$ is randomly drawn from three classes: (a) {\small XOR}, 
whose output is one (zero) if $i$ has an odd (even) number of inputs 
that are one; (b) {\small OR}, whose output is one if and only if at 
least one input is one; or (c) {\small AND}, whose output is one if 
and only if all $K_i$ inputs are one.  Following \cite{shmulevich04}, 
we refer to {\small XOR} as highly sensitive because \textit{any} 
single input flip will cause its output to flip, so $R_{ij} = 1$ 
whenever there is an edge from $j$ to $i$.  On the other hand, if 
node $i$ has {\small OR} or {\small AND} as its update rule, flipping 
node $j$ will cause node $i$ to flip only if every node other than 
$j$ is zero or one, respectively.  Thus in cases (b) and (c), $R_{ij}$ 
depends on the node biases $p_j$ which are obtained by solving 
Eqs.~(\ref{eq.p}--\ref{eq.PrX}).  For {\small OR}, $R_{ij} = \prod_k 
(1-p_k)$, and for {\small AND}, $R_{ij} = \prod_k p_k$, where the 
products are taken over all inputs $k$ which are not equal to $j$. 

Figure~\ref{fig}($a$--$c$) shows results for a network with these 
three classes of update rules.  We assign a fraction of nodes $\alpha$ 
to have {\small XOR} update rules, and the remaining nodes are evenly 
split between {\small OR} and {\small AND} rules.  We consider three 
cases: (1) {\small XOR} is assigned to the $\alpha N$ nodes with the 
largest in-degree; (2) {\small XOR} is randomly assigned to $\alpha N$ 
nodes irrespective of their degrees; or (3) {\small XOR} is assigned 
to the $\alpha N$ nodes with smallest in-degree.  In all three cases, 
the remainder are randomly assigned {\small OR} or {\small AND}.  In 
numerical simulations, all update rule assignments are quenched.  In 
order to find appropriate semi-annealing probabilities to use in our 
theoretical prediction, we note that the initial assignment of {\small 
XOR} is deterministic in cases (1) and (3), but {\small OR} and {\small 
AND} are assigned randomly.  Therefore, in the theory, we treat the 
network and the identity of the {\small XOR} nodes as fixed, and anneal 
over the {\small OR} and {\small AND} nodes, assigning a probability 
of $\nicefrac{1}{2}$ to choosing either {\small OR} or {\small AND} 
on each time step. (Other annealing choices are also possible, but 
we choose this because it most straightforwardly resembles the quenched 
assignment of update rules above.)  

As can be seen in Fig.~\ref{fig}($a$), the values of $\alpha$ at which 
the three cases become unstable are quite different, thus demonstrating 
that stability is strongly affected by correlation between the local 
topological property of nodal in-degree and the sensitive {\small XOR} 
update rule.  As shown in Fig.~\ref{fig}($c$), however, when we re-plot 
$\langle H \rangle$ against $\lambda_R$, we see that in each case the 
network becomes unstable at $\lambda_R \approx 1$, as predicted by the 
theory.  This is also strikingly illustrated by the vertical arrows in 
Fig.~\ref{fig}($a$) marking the values of $\alpha$ at which $\lambda_R=1$ 
[c.f., Fig.~\ref{fig}($b$)].

\subsection*{Example 2: Threshold networks} 
\indent \indent
We now consider networks with threshold rules as given in Eq.~% 
\eqref{eq.threshold}; such threshold rules may be re-cast as 
Boolean functions $F_i$ by enumerating all possible $X_i$ 
and calculating whether the weighted sum of inputs exceeds 
the threshold $\theta_i$.  Conversely, threshold rules are 
appropriate for Boolean network applications in which each 
edge has a fixed ``activating'' or ``repressing'' character. 

Our results for threshold networks are shown in Fig.~\ref{fig}% 
($d$--$f$) and are generated in the following manner.  To assign 
the weight $w_{ij}$ for each edge $j \to i$, we first assign 
half of the edges to be activating and half to be repressing.  
Then, the weight is drawn from a normal distribution with mean 
$1$ (activating) or $-1$ (repressing) and standard deviation 
$\nicefrac{1}{4}$.  We also consider two additional cases where 
the weights are either correlated or anticorrelated to a topological 
property of the network, the product of a node's in-degree and 
out-degree.  (Nodes with a high degree product play a crucial 
role in the stability of Boolean networks \cite{lee07,pomerance09,% 
ott09}.)  We generate the correlated and anticorrelated cases 
by exchanging weights between pairs of edges in the original 
(``uncorrelated'') case. Specifically, we repeat the following 
procedure.  First, we select two random edges $j_1 \to i_1$ and 
$j_2 \to i_2$ in the network.  Next, we identify the edge for 
which $i$ has a higher degree product.  Finally, in the correlated 
(anticorrelated) case, we exchange the values of the two weights 
if doing so would increase (decrease) the weight going to the node 
with the higher degree product.  We repeat this procedure $E/2$ 
times, where $E$ is the number of edges in the network, so that 
each edge is expected to be considered for one exchange. 

In this example, we model the case in which the thresholds of 
different nodes are similar, but not necessarily equal.  In the 
theory, we treat this case by annealing the thresholds $\theta_i$ 
over a gaussian distribution with a mean $\bar{\theta}$ and 
standard deviation $\sigma_\theta = \nicefrac{1}{10}$.  By 
Eq.~\eqref{eq.threshold}, 
\begin{equation} 
q_i(X_i) = \Phi{\bigg [} \frac{1}{\sigma_\theta} {\bigg (} \sum_j w_{ij} x_j - 
            \bar{\theta} {\bigg )} {\bigg ]}, 
\end{equation} 
where $\Phi(x)=(2\pi)^{-1/2} \int_{-\infty}^x \exp(-t^2/2) \text{d} t$.  
Similarly, we find that $d_i(X_i,X_i^j)=\vert q_i(X_i)-q_i(X_i^j) 
\vert$. These expressions can be used to calculate $p_i$, $R_{ij}$, 
and $\lambda_R$ using Eqs.~(\ref{eq.p}--\ref{eq.PrX},\ref{eq.r}).  
(Here, as in many cases, it is not necessarily to list the ensemble 
of update rules ${\mathcal T}_i$ explicitly, because $q_i$ and $d_i$ 
can be calculated directly.)  In our numerical simulations, we treat 
$\theta_i$ as quenched by writing $\theta_i = \bar{\theta}+\delta 
\theta_i$, where $\delta \theta_i$ is drawn from a normal distribution 
with mean $0$ and standard deviation $\sigma_\theta$.  

In Fig.~\ref{fig}($d$--$f$), we show results for both a single quenched 
set of $\delta \theta_i$ (hollow markers) as well as an average over $50$ 
quenched sets of $\delta \theta_i$ (filled markers).  In each case, single 
quenched realizations show similar behavior to the average, in agreement 
with the semi-annealing hypothesis.  More striking is the qualitative 
difference between the anticorrelated case and the two other cases.  At 
low thresholds, the anticorrelated network is stable, whereas both of 
the other cases are unstable.  As the threshold is increased, the 
anticorrelated network becomes unstable before becoming stable again 
at large thresholds. This behavior is explained in Fig.~\ref{fig}($e$), 
where we see that in all three cases, $\lambda_R$ initially increases 
with increasing $\bar{\theta}$, but it is only in the anticorrelated 
that $\lambda_R < 1$ initially.  Finally, in Fig.~\ref{fig}($f$), we 
re-plot the same data for $H$ against $\lambda_R$.  We see that in all 
three cases the stability transition clearly occurs at $\lambda_R=1$, 
confirming our analysis.

\section*{Discussion} 
\indent \indent
We have presented a general framework for predicting orbit stability 
in large, locally tree-like Boolean networks, given arbitrary network 
topology and update rules.  There are three main steps in this process: 
(1) select update rule ensembles ${\mathcal T}_i$ (rather than 
deterministic rules $F_i$), and compute $q_i$ and $d_i$; (2) calculate 
the dynamical biases $p_i$ of the each node $i$ by iterating Eqs.~% 
(\ref{eq.p}--\ref{eq.PrX}); and (3) calculate the activity matrix 
$\bm{R}$ with elements given by Eq.~\eqref{eq.r}.  The largest 
eigenvalue of this matrix, $\lambda_R$, then determines the stability 
of the system, Eq.~\eqref{eq.stability}.  As illustrated above, the 
first step requires a judicious selection of which aspects of the 
update rules should remain quenched, but is typically straightforward 
thereafter.  

As examples of the application of our general stability criterion, 
we analyzed both a pedagogical case and the case of threshold networks, 
where all update rules are assumed to be of the form of Eq.~\eqref{% 
eq.threshold}.  These results show that the stability of a Boolean 
network is strongly affected not only by the network topology and 
nodal update rules, but by correlations between the two.  Although 
previous research into the stability of Boolean networks has primarily 
focused on either topology or update rules alone, Figs.~\ref{fig}% 
($a$,$d$) show that correlations can have profound qualitative effects 
on the dynamical properties of a network.  Presumably, these aspects 
of biological networks interact strongly during evolution, and so 
joint effects in topology and update rules should be analyzed 
carefully when studying genetic or neural systems. 

\textit{Acknowledgements:} This work was funded by ARO grant W911NF-12-1-0101.

\newpage

\title{Supplemental Information \vspace{-1in}}
\author{}
\date{}
\maketitle

\begin{center}
\textbf{Abstract}\\
\end{center}
\indent \indent 
In this supplement, we explore several topics related to 
our stability condition for Boolean networks.  We show that the 
stability condition is unchanged for asynchronously updated networks, 
discuss the conditions under which our derivation is valid, analyze 
the computational complexity of our solution, and calculate the 
critical slope of the stability transition.

\section*{Asynchronous updating} 
\indent \indent
Asynchronous updates may arise in discrete state systems for several 
reasons.  For example, links may have nonuniform delays, $\delta_{ij}$, 
that model delays arising from, for example, the chemical kinetics of 
gene regulation.  In this case, the dynamics would be described by 
a modified version of Eq.~\eqref{eq.dynamics3} in which the state of node $i$ at time 
$t$ depends on the states of its inputs $j$ at times $(t-\delta_{ij})$.  
Another alternative is a model in which nodes are individually chosen 
to be updated in a stochastically determined order.  Here, we show that 
the stability condition given in the main text applies not only to the 
case of synchronous nodal updates, but to asynchronous models as well, 
including both of these examples.  

In particular, we consider update times, $\tau_1 < \tau_2 < ... < 
\tau_t < ...$, where the update intervals, $(\tau_{t+1} - \tau_t)$, 
are arbitrary, incommensurate and do not influence the analysis.  
Since the update times are incommensurate, we approximate the 
deterministic choice of node to update at each time, indexed by 
integer $t$, with a stochastic process where node $i$ (and only 
node $i$) is chosen to be updated by Eq.~\eqref{eq.dynamics3} with probability 
$\rho_i$.  This is, of course, also appropriate to systems that 
are inherently stochastic.  To analyze this case, we define the 
vector $\bm{y}(t)$ as in the main text; however, we must make 
some adjustments to the approximate update equation, Eq.~\eqref{eq.r}. 
Since node $i$ is chosen independently of the values of the nodes, 
the joint probability at time step $t$ that node $i$ is chosen for 
update and that node $i$ differs between the two initial conditions 
after the update is given approximately by $\rho_i \sum R_{ij} y_j(t)$.  
If node $i$ is not chosen for update at this time step, $y_i$ does 
not change.  Putting this together, we get, for small $t$ and small 
initial perturbations, $y_i(t+1) \approx \rho_i \sum R_{ij} y_j(t) 
+ (1-\rho_i) y_i(t)$, which we rewrite in matrix form as 
\begin{equation} 
\tag{S1}
\label{eq.async} 
\bm{y}(t+1) = \bm{\rho} (\bm{R} - \bm{I}) \bm{y}(t) + \bm{y}(t), \nonumber
\end{equation} 
where $\bm{\rho}$ is a diagonal matrix with $\rho_i$ in each row, 
$\bm{I}$ is the identity matrix, and $\bm{R}$ is the activity matrix.  
In order to see that Eq.~\eqref{eq.stability} also applies in this case, we note that, 
at criticality, $\bm{y}(t+1) = \bm{y}(t)$, so that Eq.~\eqref{eq.async} 
reduces to $\bm{\rho} (\bm{R} - \bm{I})\bm{y}(t) = \bm{0}$.  This has a 
solution for $\bm{y} \neq \bm{0}$ only if $\lambda_R = 1$.  Note, however, 
that in this case, for $\lambda_R > 1$, the growth rate of the Hamming 
distance will be at a rate of the order of $1/N$ smaller than the rate 
of the synchronously updated networks, because $N$ time steps of 
asynchronous update correspond to one time step of synchronous update.

\section*{Comments on Equations~(5--6)}
\indent \indent
In the main text, we make three simplifying assumptions about the 
derivation and solution of Eqs.~(\ref{eq.p}--\ref{eq.PrX}):  
\begin{enumerate}
\item The correlations between the states of different inputs to a 
      single node are negligible, because the network is locally treelike.  
\item Equations~(5--6) have a single stable solution, describing 
      the attractor of the semi-annealed dynamics.  
\item This solution can be found by iterating the equations. 
\end{enumerate}
Here we add some comments about these assumptions and the conditions 
under which they are valid.

First, the locally treelike approximation has been effective in 
describing the structural and dynamical properties of a variety of 
complex networks, as documented in \cite{melnik11}.  In particular, 
it has been applied successfully to Boolean networks and related 
percolation problems in \cite{restrepo08,pomerance09,pomerance12,% 
squires12}.  In this context, we may argue as follows that it allows 
us to assume the independence of two nodes $j_1$ and $j_2$ which are 
both inputs to a third node $i$.  We would expect correlations between 
$x_{j_1}(t)$ and $x_{j_2}(t)$ to arise mainly from the two nodes being 
mutually influenced by a previous state of a third node, $x_k(t-h)$, 
where node $k$ has paths of length $h$ to both $j_1$ and $j_2$.  But 
if this is the case, $k$ has two independent paths of length $h+1$ to 
$i$.  In locally treelike networks, the number of such paths is an 
insignificant fraction of paths of the same length.  It is hypothetically 
possible that this assumption may nonetheless break down in cases where 
the dynamics exhibit a long correlation length, as might be expected 
when there is a phase transition in $\langle p_i \rangle$.  Numerically, 
however, we find no cause for concern.  For example, such a case occurs 
for the threshold networks studied in Example 2 of the main text, when 
the stability transition at large $\bar{\theta}$ coincides with a phase 
transition in $\langle p_i \rangle$.  In this case, as in all of our 
numerical work, we observe that the stability transition still occurs 
at $\lambda_R=1$, as predicted by theory. 

We next consider the second and third assumptions above.  The Brouwer 
fixed point theorem guarantees that Eqs.~(\ref{eq.p}--\ref{eq.PrX}) have at least one 
solution, but not that it is unique or stable.  Non-uniqueness does 
not present any difficulties for the theory; in this case, each 
solution represents a separate attractor, and the stability of each 
attractor may be determined separately.  (For example, it is possible 
to construct threshold networks which have one solution with $p_i=0$ 
for all nodes $i$ and another solution where $p_i>0$ for most $i$.)  
A second, more troublesome scenario is that there are no stable 
solutions.  In this case, iteration of Eqs.~(\ref{eq.p}--\ref{eq.PrX}) would not converge 
to a solution but instead fluctuate periodically or chaotically.  An 
example of this behavior occurs in networks where the update rules 
are chosen to approximate the logistic map.  In this case, as the 
tuning parameter is changed, the system undergoes a period-doubling 
cascade.  In principle the method could be extended to this situation; 
however, it seems unlikely that systems undergoing significant dynamics 
in the biases would be stable with respect to small perturbations.  
We note, however, that this is a rather pathological case, and that 
typical biological applications have stable solutions.  

One final possibility is that there is a family of marginally stable 
solutions to Eqs.~(\ref{eq.p}--\ref{eq.PrX}).  For example, this occurs in a loop with 
two nodes and the the copy update function (i.e., the output is the 
input).  In this case, iteration oscillates and does not converge, 
since any solution where $p_1=p_2$ is valid.  We have never observed 
this phenomenon when $0<q_i(X_i)<1$ for all $i$ and $X_i$, but this
sometimes occurs when Eqs.~(\ref{eq.p}--\ref{eq.PrX}) are applied directly to quenched, 
deterministic dynamics (i.e., $q_i(X_i) = 0$ or $1$ for all $i$ and 
$X_i$).  However, when analyzing deterministic dynamics, one may 
either consider a related semi-annealed problem (that reflects, 
say, realistic noise models or measurement uncertainty), as we do 
here, or one may measure $\Pr[X_i]$ for a particular attractor 
directly from numerical simulations, then find the stability 
condition using Eqs.~\eqref{eq.r} and \eqref{eq.stability}.

\section*{Computational Complexity}
\indent \indent
We note that the procedure presented in the main text is applicable 
even to very large networks.  The Frobenius-Perron eigenvalue $\lambda_R$ 
may typically be found through power iteration, which requires only 
$\mathcal{O}(E)$ operations, where $E$ is the number of edges; for 
sparse networks, this is $\mathcal{O}(N)$.  Another advantage of our 
method is that the use of dynamical biases and the locally treelike 
approximation offers a tremendous computational improvement over 
previous theoretical treatments of similar systems.  For example, 
the analysis of probabilistic Boolean networks in Ref.~\cite{% 
shmulevich02} relies upon a state transition matrix of size $2^N 
\times 2^N$, which is intractable for networks with more than a 
few dozen nodes.  In contrast, iterating Eqs.~(\ref{eq.p}--\ref{eq.PrX}) requires 
fewer than ${\mathcal O}(N 2^K)$ steps, where $K$ is the maximum 
in-degree of all nodes.  This is numerically feasible for large-$N$ 
networks as long as $K \le 20$.  In many cases, additional 
simplifying assumptions may offer even greater computational speed.

\section*{Critical Slope} 
\indent \indent
The second-order terms in the expansion in Eq.~\eqref{eq.y2} may be used to 
derive the critical slope of $H$ near $\lambda_R=1$.  We include a 
sketch of this derivation because it may be useful for near-critical 
approximations or for designing networks with extreme behavior near 
the critical point.  

To find the second-order terms in Eq.~\eqref{eq.y2}, we need to consider 
input combinations which differ for two distinct inputs $j$ and 
$k$, which we denote $\tilde{X}=X_i^{j,k}$ in analogy with our 
definition of $X_i^j$ in the main text.  The probabilities of 
these input combinations are given, up to ${\mathcal O}(y^2)$, by 
\begin{equation} 
\tag{S2}
\begin{split} 
\Pr\left[ X_i(t), X_i^j(t) \right] &= \Pr[X_i] \ y_j(t) \left( 1 - \sum_{k 
                                      \ne j} y_k(t) \right) \\ 
\Pr\left[ X_i(t), X_i^{j,k}(t) \right] &= \Pr[X_i] \ y_j(t) \ y_k(t). 
\end{split} 
\end{equation} 
Following similar steps as those that led to Eq.~\eqref{eq.yt}, we obtain 
\begin{equation} 
\tag{S3}
\begin{split} 
y_i(t+1) &= \sum_j R_{ij} y_j(t) + \sum_{j,k} R_{ijk} y_j(t) y_k(t), \\ 
R_{ijk} &\equiv \frac{1}{2} \sum_{X_i} \Pr[X_i] d_i\left[ X_i, X_i^{j,k} 
                \right] - R_{ij}, 
\end{split} 
\end{equation} 
where $R_{ij}$ is defined as in the main text.  Note that when $j=k$, 
$R_{ijk}=0$. 

Now we may derive the critical slope.  We consider each $y_i$ to 
have reached a steady state and hence drop the time-dependence in 
$y_i(t)$.  Next we write a perturbation expansion for each variable 
near the critical point, $y_i = \varepsilon H y_i^1 + \varepsilon^2 
y_i^2$ and $\lambda_R = 1 + \varepsilon \lambda_R^1$, where superscripts 
for $y_i$ and $\lambda_R$ refer to the level of the perturbation 
expansion.  From Eq.~\eqref{eq.yt}, $\bm{y}^1$ must be the right Frobenius-% 
Perron eigenvector of the first-order $R$-matrix.  Here it has been 
normalized so that $\sum_i y_i^1 = 1$.  Inserting the second-order 
expansion and simplifying, we obtain  
\begin{equation} 
\tag{S4}
\label{eq.derivation} 
y_i^2 = H \lambda_R^1 y_i^1 + \sum_j R_{ij} y_j^2 + H^2 \sum_{j,k} R_{ijk} y_j^1 y_k^1 
\end{equation} 
This expression may be further simplified by using left Frobenius-Perron 
eigenvector of $R$, which we denote $\bm{u}$.  Multiplying through by 
$u_i$ and summing over $i$, the left-hand side and the second term on 
the right-hand side of Eq.~\eqref{eq.derivation} cancel to leading order 
in $\varepsilon$.  With the remaining terms, we find that the critical 
slope $m_c = H / \lambda_R^1$ is 
\begin{equation} 
\tag{S5}
m_c = -\frac{\sum_i u_i y^1_i}{\sum_{i,j,k} R_{ijk} u_i y_j^1 y_k^1}. 
\end{equation} 
This result may be used numerically to find the critical slope in 
particular cases, because the eigenvector $\bm{y}^1$ may be found 
along with $\lambda_R$.  It may also be used to approximate the 
critical slope analytically, when good approximations for $\bm{y}^1$ 
are known, as in Refs.~\cite{pomerance09,ott09}.

\bibliographystyle{unsrt}
\bibliography{bib}

\end{document}